# HIGH SPEED MULTIPLE VALUED LOGIC FULL ADDER USING CARBON NANO TUBE FIELD EFFECT TRANSISTOR

Ashkan Khatir[1] , Shaghayegh Abdolahzadegan[2] ,Iman Mahmoudi

Islamic Azad University,Science and Research Branch, Tehran, Iran

[1]a.khatir@srbiau.ac.ir
[2]s.abdolahzadegan@srbiau.ac.ir

*Abstract*

*High speed Full-Adder (FA) module is a critical element in designing high performance arithmetic circuits. In this paper, we propose a new high speed multiple-valued logic FA module. The proposed FA is constructed by 14 transistors and 3 capacitors, using carbon nano-tube field effect transistor (CNFET) technology. Furthermore, our proposed technique has been examined in different voltages (i.e., 0.65v and 0.9v). The observed results reveal power consumption and power delay product (PDP) improvements compared to existing FA counterparts.*

## 1. INTRODUCTION

Over the past years, silicon-based electronic technology has been improved through downscaling MOSFETs, resulting in higher device performance and density. However, there are still some obstacles to scaling, such as diffusion areas will no longer be separated by a low doped channel region and equivalent gate oxide thickness will fall below the tunnelling limit. Hence, it is important to extend or complement traditional silicon technology. As one of the promising technologies, nanotechnology avoids most of the fundamental limitations for conventional silicon devices. Nanoelectronic is an applicable field of nanotechnology that is producing nanoscale machines and systems efficiently, such as nanowires, nanoparticles and Carbon Nano Tubes (CNT). CNTs have special electronic, thermal and mechanical properties that make them attractive for the future integrated circuit applications. Transistors with carbon nanotubes as their channel are called Carbon Nanotube Field Effect Transistor (CNFET). Recently, some circuit applications are presented based on CNFETs, such as ring oscillators, invertors, and logic gates [1]. Arithmetic operations are extensively used in many VLSI applications such as signal processing, and digital communications [2], [3]. Adders are major part of computational circuits that are used for implementing any other arithmetic operation such as subtraction, multiplication or even logarithmic functions [4],[5]. Hence, efficiency of Adders affects performance of whole system therefore designers attempt to produce more efficient Adders. Many logic styles have been designed since now to produce efficient Full Adder cells. The complementary CMOS and CPL designs are two conventional Adders based on CMOS structure. Based on transmission function and transmission gate, TFA and TGA designs were implemented. The other designs are classified as Hybrid designs. Applying

CNFETs in Full Adder cells will improve delay and power consumption of circuit. In this paper a novel CNFET high speed Full Adder cell is proposed.

## 2. CARBON NANOTUBE FIELD EFFECT TRANSISTORS (CNFETS)

Carbon Nano Tubes (CNTs) are sheets of graphene rolled into tube [4]. Depending on their chirality (i.e., the direction in which the graphite sheet is rolled), the single-walled carbon nanotubes can either be metallic or semiconducting [5], [6]. CNFETs are one of the molecular devices that avoid most fundamental silicon transistor restriction and have ballistic or near ballistic transport in their channel [7], [8]. Therefore a semiconductor carbon nanotube is appropriate for using as channel of field effect transistors [6]. Applied voltage to the gate can control the electrical conductance of the CNT by changing electron density in the channel.

By using appropriate diameter suitable threshold voltage for CNFET can be achieved. The threshold voltage of the CNFET is proportional to the inverse of the diameter of CNT and can be expressed as:

$$V_{TH} = \frac{0.42}{d(nm)} \tag{1}$$

For a CNT with (n, m) as chirality and a as lattice (that is carbon to carbon atom distance) the diameter is [10]:

$$D_{CNT} = \frac{a\sqrt{n^2 + nm + m^2}}{\pi} \tag{2}$$

As mentioned above, CNTs are used in CNFETs as channel and depending on the connections between source and drain with channel (CNTs) there are two main CNFETs. It works on the principle of direct tunnelling through a Schottky barrier at the source–channel junction [11]; therefore, these transistors are called Schottky Barrier CNFET (SB-CNFET). SB-CNFET shows ambipolar behavior and Constrain usage of these transistors in conventional CMOS-like logic families. Schottky barrier restricts the transconductance in the ON state, and thus Ion/Ioff ratio becomes rather low [12]. Second device is MOSFET-like CNFET which is doped in un-gated portions and has similar behaviour to CMOS transistors and it presents unipolar behaviour [13]. The semiconductor-semiconductor junction will eliminate schottky barrier and therefore there is higher ON current unlike SB-CNFETs. Other advantages of MOSFET-like CNFETs are high on-off ratio and their scalability compared to their schottky barrier counterparts [8].  In this paper we utilized MOSFET-like CNFETs for proposed design.

## 3. MULTIPLE VALUED LOGIC

Multiple valued logic is an extension to classical two-valued logic as n-valued logic for n>2. MVL circuits can reduce the number of operations necessary to implement a particular mathematical function in addition to have an advantage in terms of reduced area [14]. These reductions can improve total circuit performance.

## 4. PREVIOUS WORKS

Rapid improvements of technology drive designers to compete for lower power consumption, smaller silicon area, higher speed and higher throughput. One of the major parts of every processor, which participate in arithmetic logic unit floating point unit and address generation, is Full Adder [14-16]. Since, the central part of digital computing lies in FA cell, increasing the performance of this part can improve total performance dramatically.

There are two design styles for Full Adder cells. In classical Full Adder designs only one logic style is used for the whole Full Adder cell, whereas hybrid Full Adders use more than one logic style in a cell [17]. The static complementary CMOS Full-Adder is a 28 transistor Full Adder Logic style, which composed of pull-up PMOS and pull-down NMOS networks. C-CMOS provides full swing output and good driving capabilities. The main drawback of C-CMOS is transistor count that causes large chip area and large input capacitance [18]. Another conventional logic is complementary pass transistor logic (CPL). In CPL two different parts are used for implementing SUM and Carry outputs. The logic style is a high speed and full swing design, but due to the presence of a lot of internal nodes and static inverters, there is large power dissipation [19]. Number of used transistors in this logic is 32 that results in complication in the layout. Transmission gate Full Adder (TGA) is introduced for its low power dissipation but it lacks driving capability. All of the above mentioned Full Adders are using single logic style. Hybrid CMOS Full Adder is composed of three modules First module produces intermediate signals that pass to second and third modules and it can be XOR-XOR, XOR-XNOR, XNOR-XNOR [17],[20],[21] or even majority-not function. Second module produces a SUM and third one Produces Carry. One example of this logic style Adder is Static Energy Recovery Full Adder (SERF) that uses 10 transistors but its output is not full swing and has not a good driving capability for next modules in cascading modes. NEW14T Adder [22], DB cell [17], SERF Adder [20] and HPSC [24] are examples of hybrid Full Adders. Some circuits are designed by majority function.

Majority function is a logic circuit that performs as a majority vote to determine the output of the circuit. This function has odd number of inputs and thus it operates as Carry-out in Full Adder. The majority function is represented as (3):

$$\text{Majority}(A, B, C) = AB + AC + BC = C_{out} \qquad (3)$$

In this structure majority-not gate is created with input capacitors and a static CMOS inverter at follow. In CMOS Technology NAND, NOR and majority-not gates can be implemented by scaled Inverters [25]. In order to implement them by CNT transistors the diameter of CNFETs must be adjusted.

Different design of majority-not based Full Adders reported in [25-28]. These circuits are implemented in two stages as shown in Fig 1. Although for producing carry out they all have similar function, but in SUM output they differ in circuit structure and the transistor count. The first stage is producing $\overline{C_{out}}$ by means of majority-not function and in the second stage SUM is generated by means of functions F that implemented in different styles. These Full Adders have a simple structure but it operates very well and results in remarkable advances in reducing power in comparison to other well-known designs. This reduction is due to simple structure, reduced number of transistors and the lowering in switching activities. [25]

First majority-not based Full Adders is implemented by function of

$$\text{SUM} = \text{Maj}(A, B, \overline{C_{out}}, \overline{C_{out}})[25] \qquad (4)$$

In this design two stages majority-not function is used first one for $\overline{C_{out}}$ and next stage as five input majority-not function for creating Sum. The logic of [25] is composed of two parallel connected NMOS and PMOS transistors that provide a path to the input. Transistor count is 8 and capacitor count is 7.

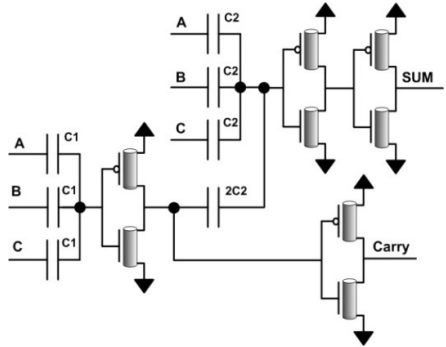

Figure 1.  Full Adder Design CNT-FA1 [25]

The another design low-voltage CNFET-based Full Adder circuit based on Minority Function presented in [26] uses only 8 transistors and reduces the number of capacitors to 5 as shown in fig 2, but uses capacitors in the middle of design which result in reduction in speed. The $\overline{C_{out}}$ implemented by majority-not function too.

In next design [27], NAND and NOR gates are used for implementing Sum as equation of $SUM = Maj(A, B, 2*Nand(A, B, C), 2*Nor(A, B, C))$. By reducing the number of capacitors, overall performance of the circuit is improved. Figure 3

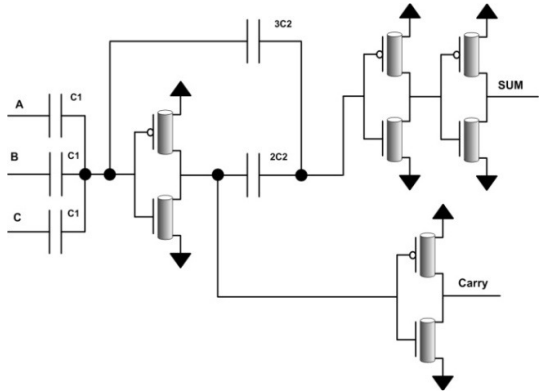

Figure2.Full Adder Design CNT-FA3 [26]

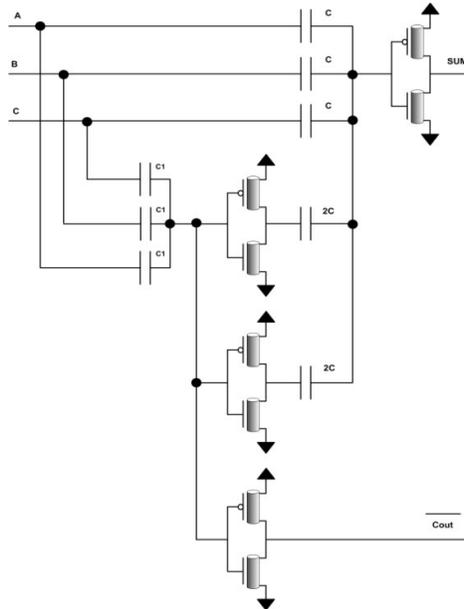

Figure 3. Full Adder Design CNT-FA2 [27]

## 5. PROPOSED FULL ADDER

One method to implement majority-not function is using a conventional inverter and three capacitors connected to the inputs of inverter as presented in Fig 4. In design of majority-not function, three capacitors are implementing Multiple-Valued logic by producing four logic levels. The levels are $0, \frac{V_{DD}}{3}, \frac{2V_{DD}}{3}$ and $V_{DD}$. The inverter threshold voltage is regulated so that it will be OFF when most number of inputs are '1'. By changing threshold voltage of inverter transistors of Fig.4, the circuit can be used as Either NAND or NOR functions.

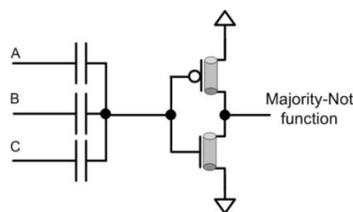

Figure 4. Majority-not Function Circuit

For NAND function, when three inputs (A, B, C) are all '1', output should be '0' otherwise output should be '1'. For NOR function if there is one input that is '1', output should be '0' otherwise output should be '1'. As in sum formula (4) output can implemented by NOR, NAND and majority-not functions and Fig. 5 and formula (5) will be attained.

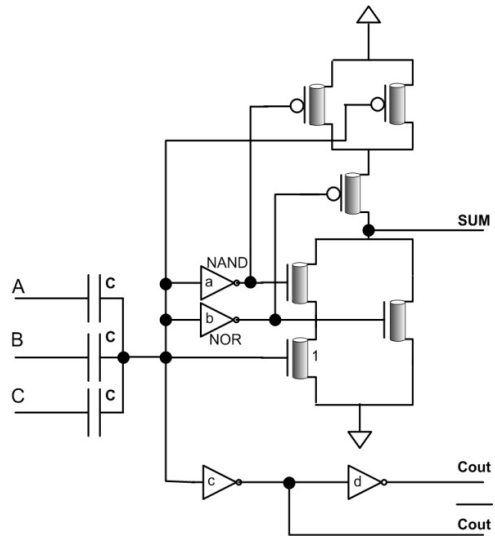

Figure 5.  Proposed Full Adder design

(5)

In proposed Adder we use less capacitors rather than contemporary CNFET Adders. The simulation results indicate that this reduction has caused our Full Adder cell to become much faster. This design presents a circuit which uses 14 transistors and 3 capacitors and also produces carry and          outputs. Outputs of the circuit will be connected to power supply or ground and therewith, the circuit has good drive capability. In this design, "a" and" b" inverters implement NOR and NAND functions respectively. When at least one input is '1', the transistor marked with '1' will act as a majority function and becomes ON. The proposed design is full swing and in comparison to previous designs uses fewer capacitors.

## 6. SIMULATION RESULTS

The Synopsys HSPICE circuit simulator has been used to simulate Full Adder circuits. The circuits of C-CMOS, CPL, TGA and Hybrid are simulated using 32nm MOSFET technology. For simulating CNFET-based circuits the proposed compact model in [8] is used.  This standard model has been designed for unipolar, MOSFET-like CNFET devices, in which each transistor may have one or more CNTs. Circuits are compared by power consumption, propagation delay and power delay product (PDP). For being more realistic, buffers (two cascaded inverter) are placed at the outputs. The proposed design is compared with MOSFET designs and earlier CNFET Full Adder designs CNT-FA1 [25] CNT-FA2 [27] and CNT-FA3 [26].  Table 1 and 2 shows the value of power, delay and PDP for C-CMOS, CPL, TGA, Hybrid, CNT-FA1, CNT-FA2 and CNT-FA3. CNFETs are simulated in 0.9v and 0.65v. Delay is measured from middle of the input voltage swing to the middle of the output voltage swing. Simulation results in Table 1 and 2, indicate that the smallest delay belongs to the proposed Full Adder cell. PDP is calculated as a trade-off between power consumption and delay. It is a measure of total performance of a circuit. From Table 1 and 2, it can be seen that the proposed Full Adder has the best PDP.

TABLE 1.        Simulation Results for 0.9v

| Design | Performance Parameters | | |
|--------|-------|-------|-----|
|  | *Power* | *Delay* | *PDP* |
| C-CMOS | 6.26E-07 | 5.27E-11 | 3.30E-17 |
| CPL | 4.87E-07 | 1.57E-10 | 7.63E-17 |
| TFA | 6.32E-07 | 2.89E-10 | 1.83E-16 |
| TGA | 6.68E-07 | 3.46E-10 | 2.31E-16 |
| Hybrid | 4.96E-07 | 2.93E-10 | 1.45E-16 |
| CNT-FA1 | 1.05E-06 | 7.83E-11 | 8.20E-17 |
| CNT-FA2 | 3.32E-07 | 1.14E-10 | 3.80E-17 |
| CNT-FA3 | 7.83E-07 | 5.36E-11 | 4.20E-17 |
| Proposed Adder | **8.40E-07** | **1.25E-11** | **6.17E-18** |

TABLE 2.        Simulation Results for 0.65v

| Design | Performance Parameters | | |
|--------|-------|-------|-----|
|  | *Power* | *Delay* | *PDP* |
| C-CMOS | 2.94E-07 | 1.46E-10 | 4.28E-17 |
| CPL | 2.08E-07 | 4.65E-10 | 9.67E-17 |
| TFA | 1.52E-07 | 8.45E-10 | 1.29E-16 |
| TGA | 1.21E-07 | 4.76E-10 | 5.77E-17 |
| Hybrid | 1.71E-07 | 1.10E-09 | 1.88E-16 |
| CNT-FA1 | 5.23E-07 | 7.97E-11 | 4.17E-17 |
| CNT-FA2 | 4.71E-07 | 8.82E-11 | 4.15E-17 |
| CNT-FA3 | 7.12E-07 | 7.51E-11 | 5.35E-17 |
| Proposed Adder | **4.80E-07** | **1.29E-11** | **1.05E-17** |

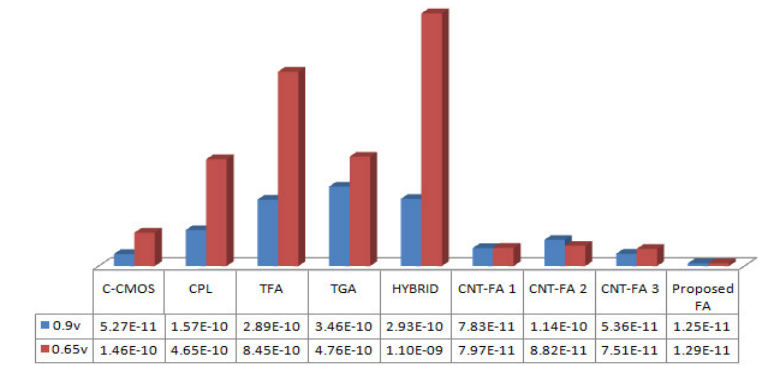

| | C-CMOS | CPL | TFA | TGA | HYBRID | CNT-FA 1 | CNT-FA 2 | CNT-FA 3 | Proposed FA |
|---|--------|-----|-----|-----|--------|----------|----------|----------|-------------|
| ■ 0.9v | 5.27E-11 | 1.57E-10 | 2.89E-10 | 3.46E-10 | 2.93E-10 | 7.83E-11 | 1.14E-10 | 5.36E-11 | 1.25E-11 |
| ■ 0.65v | 1.46E-10 | 4.65E-10 | 8.45E-10 | 4.76E-10 | 1.10E-09 | 7.97E-11 | 8.82E-11 | 7.51E-11 | 1.29E-11 |

Figure 6. Propagation Delay comparison chart for 0.9v and 0.65v

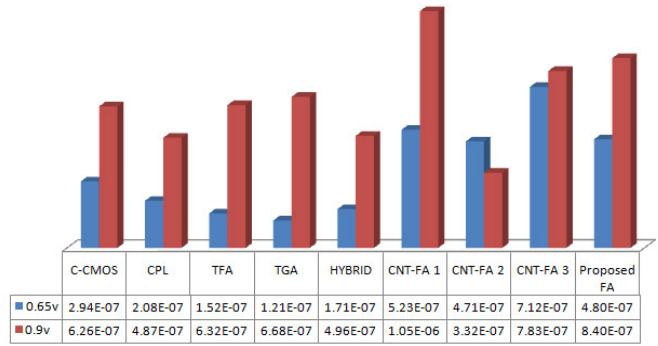

| | C-CMOS | CPL | TFA | TGA | HYBRID | CNT-FA 1 | CNT-FA 2 | CNT-FA 3 | Proposed FA |
|---|---|---|---|---|---|---|---|---|---|
| ■ 0.65v | 2.94E-07 | 2.08E-07 | 1.52E-07 | 1.21E-07 | 1.71E-07 | 5.23E-07 | 4.71E-07 | 7.12E-07 | 4.80E-07 |
| ■ 0.9v | 6.26E-07 | 4.87E-07 | 6.32E-07 | 6.68E-07 | 4.96E-07 | 1.05E-06 | 3.32E-07 | 7.83E-07 | 8.40E-07 |

Figure 7. Power consumption comparison chart for 0.9v and 0.65v

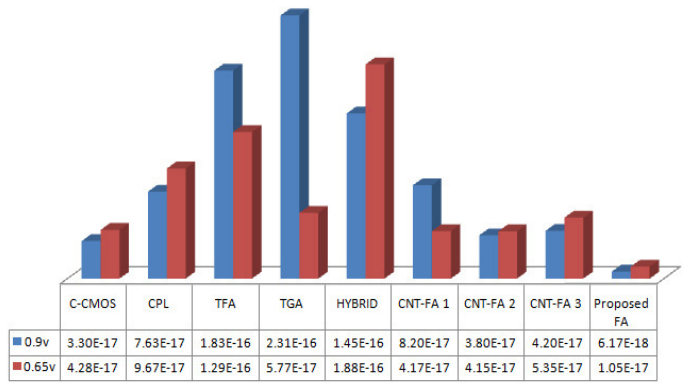

| | C-CMOS | CPL | TFA | TGA | HYBRID | CNT-FA 1 | CNT-FA 2 | CNT-FA 3 | Proposed FA |
|---|---|---|---|---|---|---|---|---|---|
| ■ 0.9v | 3.30E-17 | 7.63E-17 | 1.83E-16 | 2.31E-16 | 1.45E-16 | 8.20E-17 | 3.80E-17 | 4.20E-17 | 6.17E-18 |
| ■ 0.65v | 4.28E-17 | 9.67E-17 | 1.29E-16 | 5.77E-17 | 1.88E-16 | 4.17E-17 | 4.15E-17 | 5.35E-17 | 1.05E-17 |

Figure 8 PDP comparison chart for 0.9 and 0.65v

## 3. CONCLUSIONS

In this paper a novel high speed majority based CNFET Full Adder has been proposed. Using CNFETs as a novel Full-Adder architecture improved the efficiency. The main concept of this design is implementing (5) using majority-not function and using majority-not circuit design as NOR and NAND functions by changing threshold voltage of CNFETs to produce intermediate signals. In order to evaluate the performance of the design, delay, power and power-delay-product (PDP) factors are compared with some of the state-of-the-art MOS and CNFET-based designs. Simulations have been performed on HSPICE by using CNFET technology in two practical voltages as 0.9v and 0.65v. To evaluate new design eight other designs, including CMOS, CPL, TFA, TGA, Hybrid and three previous majority not based Full Adder designs are applied. Simulation results illustrate improvements in terms of delay and PDP in comparison to previous MOSFET and CNFET designs. The improvement in terms of PDP obtained by the proposed Full-Adder cell as compared the best standard design amounts to 18% in 0.9v and 36% in 0.65v.

## ACKNOWLEDGEMENTS

The authors would like to thank Dr. Belmond Yoberd for his literature contribution.